\documentclass{elsart}

 \usepackage{epsfig}

\usepackage{amssymb}

\begin{document}

\begin{frontmatter}



\title{Antiprotonic Potentials from Global Fits to 
the PS209 Data\thanksref{titlefn}}
\thanks[titlefn]{Supported in part by the Israel Science Foundation}

\author{E. Friedman\thanksref{cor}\thanksref{ead}},
\author{A. Gal}
\thanks[cor]{Corresponding author.}
\thanks[ead]{E mail address: elifried@vms.huji.ac.il}

\address{Racah Institute of Physics, The Hebrew University, Jerusalem 91904,
Israel}

\begin{abstract}

The experimental results for strong interaction effects in antiprotonic
atoms by the PS209 collaboration consist of  high quality data for 
several sequences of isotopes
along the periodic table.
Global analysis of these data in terms of a $\bar p$-nucleus optical
potential achieves good description of the data using  a $s$-wave
finite-range $\bar p N$  interaction. Equally good fits are also obtained
with a poorly-defined zero-range potential containing a $p$-wave term. 
\end{abstract}

\end{frontmatter}

\section{Introduction}
\label{sect:intro}
In  earlier publications \cite{BFG95,BFG97} most of the then available data
on strong interaction effects in antiprotonic atoms have been used in
`global' analyses in terms of a $\bar p$-nucleus optical potential.
The simplest form of the potential is

\begin{equation}
\label{equ:potl}
2\mu V_{{\rm opt}}(r) = -4\pi(1+\frac{\mu}{M}
\frac{A-1}{A})[b_0(\rho_n+\rho_p)
  +b_1(\rho_n-\rho_p)]~~, 
\end{equation}
where $\mu$ is the $\bar p$-nucleus reduced mass,  
$\rho_n$ and $\rho_p$ are the neutron and proton density
distributions normalized to the number of neutrons $N$ and number
of protons $Z$, respectively, $A=N+Z$
and $M$ is the mass of the nucleon. In the zero-range `$t \rho$' approach
the parameters $b_0$ and $b_1$ are minus the $\bar p$-nucleon isoscalar
and isovector scattering lengths, respectively, otherwise these parameters 
may be regarded as  `effective'.
The fits of Ref. \cite{BFG95} were based on 48 data points which included
only two pairs of isotopes. As a typical result we quote
Re$b_0=2.51\pm0.24$ fm, Im$b_0=3.46\pm0.27$ fm, leading to $\chi ^2$/N,
the $\chi ^2$ per point, of 1.5, which represents a very good fit to the
data. Additional terms had also been cosidered in \cite{BFG95} such as
 a $p$-wave term, which led to $\chi ^2$/N
in the range of 1.2 to 1.4. Such a reduction in $\chi ^2$ is, however,
not statistically significant in these circumstances. 
Consequently only the above
zero-range $b_0$  version has since been used in several analyses of 
experimental results.

The recent publication of results for strong interaction effects in 
antiprotonic atoms by the PS209 collaboration \cite{TJC01}  
changed the experimental situation
dramatically by providing high quality data for several sequences of isotopes
along the periodic table.   Such data hold promise to pin-point isospin
effects in the potential and/or explicit surface effects such as 
the existence of a $p$-wave term in the potential. 
This is  the scope of the present work which is based on data for
five isotopes of Ca, four isotopes each of Fe and 
of Ni, two isotopes each of Zr and of Cd, four isotopes of Sn, five 
isotopes of Te and the isotope $^{208}$Pb, all from 
results of the PS209 collaboration \cite{TJC01}. 
Note that the results for Cd, Sn and
Te isotopes have been revised very recently to include effects of E2
resonance \cite{STC02,KWC02}.
In order to expand the range of the data further we have included also 
earlier results for $^{16,18}$O (see \cite{BFG95} for references to the 
original experiments.)
In section \ref{sect:dens} we discuss the choice of nuclear density
distributions $\rho _n$ and $\rho _p$, which form an essential ingredient
of the optical potential Eq. (\ref{equ:potl}). Section \ref{sect:res}
summarizes the results and presents a `recommended' set of parameters
which best describes the data.

\section{Density distributions} \label{sect:dens}
In choosing the nuclear density distributions for use in the analysis of
strong interaction effects in antiprotonic atoms, one is faced with the
fact that due to the strength of the $\bar p$-nucleus interaction and the
dominant role played by the imaginary potential, strong interaction
effects reflect the interaction with the extreme outer regions of the
nucleus where the density is well below 10\% of the central density.
It is therefore not very reliable to infer from these data quantities 
which relate to the bulk of the density such as root mean square (rms) 
radii. In what follows we therefore rely also
on information from pionic atoms \cite{FGa03} when choosing the densities,
bacause pionic atoms are sensitive to the region where the density is 50\% 
or more of the central density.
The density distribution of the protons is usually considered known as
it is obtained from the nuclear  charge distribution \cite{VJV87} by
unfolding the finite size of the charge of the proton.
The density distribution of the neutrons is usually not known to
sufficient accuracy.

\begin{figure}
\epsfig{file=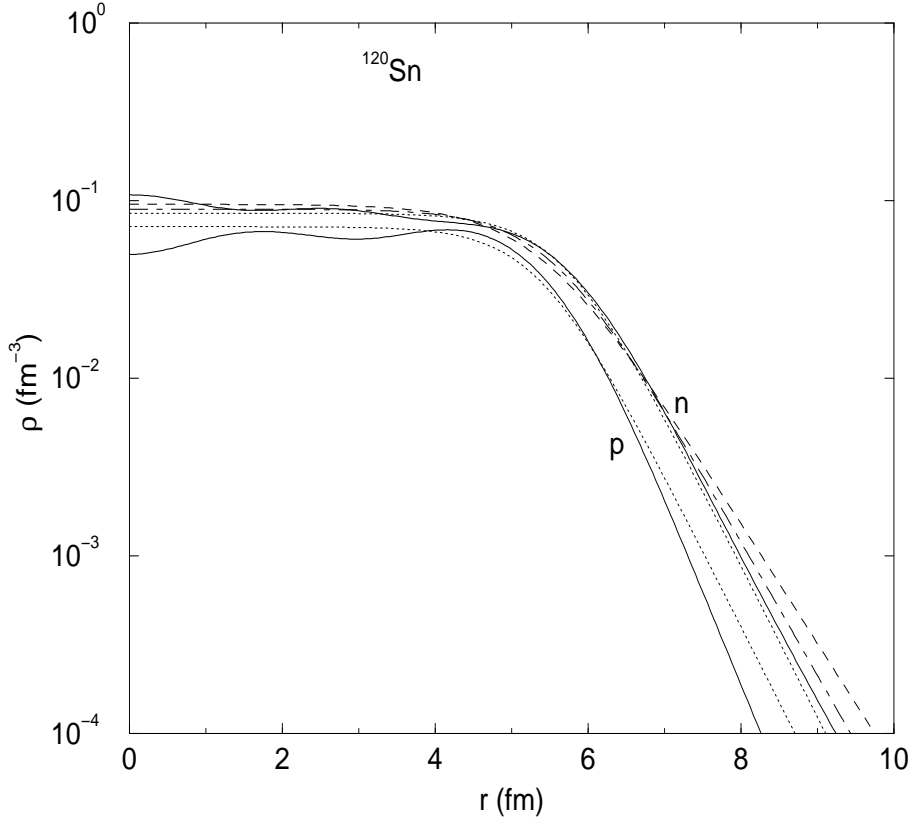, height=110mm,width=120mm}
\caption{Four neutron and two proton densities for $^{120}$Sn. 
Solid curves for SP
model, dotted curves for `skin' form, dashed curve for `halo' form and
dot-dashed curve for the `average' form. All neutron densities have the same
rms radius.}
\label{fig:dens}
\end{figure}

In a recent paper \cite{TJL01} Trzci\'nska et al. have suggested, based on 
analyses of strong interaction effects in antiprotonic atoms and of 
radiochemical data following $\bar p$ absorption on nuclei, that the difference
between neutron and proton rms radii in nuclei depend   on the
asymmetry parameter $(N-Z)/A$ as follows:

\begin{equation} \label{equ:radii}
r_n-r_p=-(0.04\pm0.03)+(1.01\pm0.15)\frac{N-Z}{A}~~{\rm (fm)}.
\end{equation}  
Addressing the question of the {\it shape} of the neutron density, they
have concluded that a `halo' rather than a `skin' shape is preferred.
These shapes are defined as follows:
 Assuming a two-parameter
Fermi distribution both for the proton (unfolded from the charge distribution)
and for the neutron density distributions

\begin{equation}
\rho_{n,p}(r)  = \frac{\rho_{0n,0p}}{1+{\rm exp}((r-R_{n,p})/a_{n,p})},
\end{equation}
then for the `skin' form
the same diffuseness parameter for the protons and the neutrons
$a_n=a_p$ is used and the $R_n$ parameter is determined from the
rms radius $r_n$.
In the `halo' form
the same radius parameter $R_n=R_p$ is assumed 
for $\rho _n$ and for $\rho _p$, and $a_n^{\rm h}$ is
determined from $r_n$. We also use an `average'
option where the diffuseness parameter is set to be the average of the
above two diffuseness parameters, namely, $a_n^{\rm ave}=(a_p+a_n^{\rm h})/2$
and the radius parameter $R_n$ is determined from the
rms radius $r_n$.
To allow for different values of $r_n-r_p$ we have used Eq.~(\ref{equ:radii})
but with variable slope parameter, where, e.g. a slope of 1.5 fm 
correspons to 
the average behaviour of results of 
relativistic mean field (RMF) calculations \cite{FGa03,LRR99}.

Figure \ref{fig:dens} shows examples of densities used in the present work.
The single-particle (SP) densities are described in \cite{BFG97}. 
The various  neutron 
densities have the 
same value of $r_n$, as given by RMF calculations.
Note that where the densities are well below 10\% of their central
values the neutron densities are  significantly larger than the proton density.

\section{Results}  \label{sect:res}
Fits were made to 90 data points using a variety of shapes of nuclear 
densities and values of $r_n$ as described in the previous section.
Finite range was also introduced as an option, where the nuclear densities
have been folded with a Gaussian,
separately for the real and for the imaginary potentials, such that the
densities in Eq. (\ref{equ:potl}) are replaced by `folded' densities
\begin{equation}
\label{equ:fold}
\rho ^F (r)~~=~~\int d{\bf r}' \rho({\bf r}') \frac{1}{\pi ^{3/2} \beta^3}
e^{-({\bf r}-{\bf r'})^2/\beta^2}~~.
\end{equation}

 Table \ref{tab:res} summarizes the main features
of the results. The columns marked  `RMF' were obtained with values
of $r_n-r_p$ taken from results of RMF calculation \cite{LRR99} whereas
the columns marked  `$\bar p$' are for $r_n-r_p$ due to Ref. \cite{TJL01}.
In both cases the 2-parameter Fermi distribution
was used  for the proton and for the neutron densities. The column
marked with SP is for both densities obtained by filling in levels in
single particle potentials \cite{BFG97} with $r_n-r_p$ values from the
RMF model. The  shapes of the neutron distributions (`ave.', 
`halo', `skin') are as explained in the previous section. 
The top part of the table is for zero range interaction and the bottom part
of the table is for finite range interaction with Gaussian range parameters
$\beta _R$ and $\beta _I$ for the real and for the imaginary parts,
respectively.
From the values of $\chi ^2$/N for the zero range fits it is seen that 
reasonable fits to the data can be achieved for all options except for the
`RMF-halo' variety. The increased values of $\chi ^2$/N compared to 
results of 
analyses of the previous data  \cite{BFG95,BFG97} 
reflect the improved accuracy and the additional
information contained in the new data. The latter is observed in  the
lower part of the table where the need
for a finite-range version of the
$\bar p$-nucleus optical potential is self-evident. 
The values of $\chi ^2$ and of
$\chi ^2$/N mean significant improvements in the fits compared to the 
zero-range results. The best values for $\chi ^2$/N are close to what
has been achieved in global fits to pionic atom data \cite{FGa03}.

\begin{table}
\caption{Global fits to 90 antiprotonic atom
 points. `RMF' and `$\bar p$' stand for values of $r_n-r_p$ used. Top part
is for zero-range potentials and bottom part for finite-range potentials
with range parameters $\beta _{R,I}$.}
\label{tab:res}
\begin{tabular}{c|c|ccc|ccc}
\multicolumn{2}{c}{} &
\multicolumn{3}{c}{`RMF'} &
\multicolumn{3}{c}{`$\overline{p}$ atoms'} \\
 &SP&ave.&halo&skin&ave.&halo&skin   \\ \hline
$\chi^2$&296&338&689&242&240&310&249\\
$\chi^2/$N & 3.3&3.8&7.7&2.7&2.7& 3.5&2.8 \\
Re$b_0$ (fm)&2.5(2)&2.6(2)&2.0(2)&3.4(2)&3.1(2)&2.7(2)&3.6(2) \\
Im$b_0$ (fm)&3.1(2)&2.8(2)&2.5(2)&3.1(2)&3.2(2)&3.1(2)&3.4(2) \\
  & & & & & & &  \\
$\chi^2$ &221&203&209&219&206&202&235 \\
$\chi^2/$N &2.5&2.3&2.3&2.4&2.3&2.2&2.6 \\
Re$b_0$ (fm)&0.78(5)&1.20(7)&0.68(5)&1.11(5)&1.77(9)&1.34(7)&0.71(3) \\
Im$b_0$ (fm)&1.75(6)&1.74(8)&1.10(4)&2.8(1)&2.3(1)&2.0(1)&3.5(1) \\
$\beta_R$ (fm)&1.0&0.9& 0.9&1.2&0.7&0.8&1.6  \\
$\beta_I$ (fm) &1.2& 0.9&1.3&0.6&0.7&0.8&0.6  \\ \hline
\end{tabular}
\end{table}

Next we turn to the question of whether  the 
 data 
require the presence of an isovector term $b_1$ in the potential 
Eq. (\ref{equ:potl}). Fits to the full data set of 90 points failed to
produce any reduction in $\chi ^2$ when such a term was introduced.
In a separate calculation we have been able to obtain a non-zero value for
the parameter $b_1$ when fits were made to only the data for the Sn
isotopes, leading to $\chi ^2$/N of 1.0 for these isotopes. However,
applying  that $b_1$ to the full data set resulted in a major deterioration
of the fit, and a further global fit restored the results
of Table \ref{tab:res} with $b_1$=0. This  is not surprizing since
antiprotonic atoms are sensitive to the nuclear density
in the extreme outer regions of the nucleus where the neutron density is
dominant.  Explicit isospin effects are, therefore, hard to observe.

Another attempt to improve the agreement between zero-range calculations
and experiment was made by introducing  a $p$-wave term into
the optical potential \cite{BFG95,BFG97}, a term which is active mostly
at the surface region. Starting with  the zero-range version of the
potential, the values of $\chi ^2$ could be reduced from those of the
upper part of Table \ref{tab:res} to those of the lower part, but with
$p$-wave parameters which depended critically on the choice of shape for
the neutron distribution. Including in addition finite-range folding
failed to further improve the fits.

In conclusion, global analyses of most of the data from PS209 reveal
a need to improve on the previously successful zero-range $s$-wave
isoscalar $\bar p$-nucleus optical potential. Including an isovector 
term is not required by the data nor is a $p$-wave term recommended at
our present state of knowledge, although it undoubtedly improves the fits
to the data when the zero-range form is used.
 The use of a finite-range interaction in the optical
potential is definitely required by the data. Assuming that the values of
the differences between rms radii $r_n-r_p$ are, on the average, 
between those advocated in Ref. \cite{TJL01} and those predicted by
RMF calculations \cite{LRR99}, and assuming the shape of the neutron 
distributions is between those described by the `skin' and by the `average'  
prescriptions, we then recommend the values Re$b_0$~=~1.3$\pm$0.1 fm,
Im$b_0$~=~1.9$\pm$0.1 fm, with $\beta_R$~=~$\beta_I$~=~0.85 fm
within the folding expression Eq. (\ref{equ:fold}).



\end{document}